\documentclass[a4paper]{jpconf}
\usepackage{graphicx}
\graphicspath{{pictures/}}
\usepackage{epstopdf}
\usepackage{citesort}

\begin{document}
\title{Features of non-congruent phase transition in modified Coulomb model of the binary ionic mixture}

\author{N E Stroev$^1$, I L Iosilevskiy$^2$}

\address{Joint Institute for High Temperatures of the Russian Academy of Sciences, Izhorskaya 13, Moscow 125412, Russia}
\address{Moscow Institute of Physics and Technology, Institutskiy Pereulok 9, Dolgoprudny 141700, Russia}

\ead{$^1$ nikita.stroev@phystech.edu,$^2$iosilevskiy@gmail.com}

\begin{abstract}
Non-congruent gas-liquid phase transition (NCPT) have been studied \cite{StIo-15} in modified Coulomb model of a binary ionic mixture C(+6) + O(+8) on a \textit{uniformly compressible} ideal electronic background /BIM($\sim$)/. The features of NCPT in improved version of the BIM($\sim$) model for the same mixture on background of \textit{non-ideal} electronic Fermi-gas and comparison it with the previous calculations are the subject of present study. Analytical fits \cite{PtCh} for Coulomb corrections to EoS of electronic and ionic subsystems were used in present calculations within the Gibbs--Guggenheim conditions of non-congruent phase equilibrium \cite{Acta}.Parameters of critical point-line (CPL) were calculated on the entire range of proportions of mixed ions $0<X<1$. Strong ``distillation'' effect was found for NCPT in present BIM($\sim$) model. Just similar distillation was obtained in variant of NCPT in dense nuslear matter \cite{PRC}. The absence of azeotropic compositions was revealed in studied variants of BIM($\sim$) in contrast to explicit existence of azeotropic compositions for the NCPT in chemically reacting plasmas and in astrophysical applications \cite{UO2,PRC}. 
\end{abstract}

\section {Introduction }
The phase transition can be called non-congruent (or incongruent), if transition between two coexisting phases involves a change in composition. In some physics fields this term is well-known, but in several communities it remains unexamined. It is important to consider a simple model, where this effect is significant, to understand all properties of studied effect.
The Potekhin and Chabrier equation of state \cite{PtCh} and related approximations were used as the base model, which are quite popular in the astrophysical community. In this study, calculations were made including all sorts of interparticle correlations in contrast to previous work \cite{StIo-15}, where the simplified model was used including only the ion-ion correlations to demonstrate the effects of non-congruence.
The modified model of binary ionic mixture, which describes fully ionized plasma, was chosen for the study. The main modification consists in changing the background electron properties. Incompressible ``rigid'' background of electrons was changed to uniformly compressible, which leads to a new gas-liquid like phase transition with an upper critical point. To describe the mixture properties the linear mixing rule (LMR) was used, since the presence of independent component (another sort of ions) always brings non-congruence effect.
\section {Equation of state for BIM($\sim$) }
At first we have to consider the equation of state of Potekhin and Chabrier and related approximations in case of binary ionic mixture. The Helmholtz free energy $F$ for the system can be written as the sum
$$F_\mathrm{tot} = F_\mathrm{id}^{i}+F_\mathrm{id}^{e}+F^\mathrm{ex},\eqno(1)$$
where first term denotes the ideal free energy of classical nonrelativistic ions of two sorts (the carbon and oxygen ions were taken with the corresponding charges and masses for calculation).
$$F_\mathrm{id}^{i} = N_{1}k_{B}T(ln(n_{1}\lambda_{1}^3)-1) + N_{2}k_{B}T(ln(n_{2}\lambda_{2}^3)-1),\eqno(2)$$
where $\lambda_{i}=(2\pi\hbar^{2}/m_{i}k_{B}T)^{1/2}$ is the thermal wavelength of ions. $k_{B}$ is the Boltzman constant, $T$ is the temperature of the equilibrium plasma, $\hbar$ is the Planck constant, $m_{i}, N_{i}, n_{i}$ are the mass, number and concentration of the ions respectively.
For electron we use
$$F_\mathrm{id}^e=N_{e}\mu_\mathrm{id}^e-P_\mathrm{id}^e V,\eqno(3)$$
where $\mu_\mathrm{id}^e$ is the electrochemical potemtial and $P_\mathrm{id}^e$ is the pressure of the ideal Fermi gas. The general model does not take into account relativistic effects, because the parameter $\tau = (k_{B}T/m_{e}c^2)$, where $m_{e}$ is the electron mass and $c$ is the speed of light can be written as $T/5.93\times 10^9 K$ and the maximum temperature of the calculations is $4.27\times 10^5 K$.
Thus, we can write
$$P_\mathrm{id}^e=\frac{(2m_{e})^{3/2}I_{3/2}(\chi)}{(3\pi^2\hbar^3\beta^{5/2})},\eqno(4)$$
$$n_{e}=\frac{\sqrt{2}(m_{e}/\beta)^{3/2}I_{1/2}(\chi)}{\pi^2\hbar^3},\eqno(5)$$
where $\beta = (k_{B}T)^{-1}$, $\chi = \beta \mu_\mathrm{id}^e$ and
$$I_{\nu}(\chi) = \int_0^\infty \frac{x^\nu}{\exp(x-\chi)+1} dx \eqno(6)$$ is the Fermi-Dirac integral.
The electrochemical potential is obtained from the relationship 
$$\chi = X_{1/2}(2\theta^{-3/2}/3),\eqno(7) $$ where $X_{\nu}$ is the inverse Fermi integral.
We use approximations that have high accuracy for these and reverse integrals from work \cite{Antia}.
The strict condition of electroneutrality is satisfied $n_{e} = Z_{1}n_{1}+Z_{2}n_{2}$, where $Z_{1}=6$ and $Z_{2}=8$.

The excess part of free energy (1) can be written as $F^\mathrm{ex}=F_\mathrm{ee}+F_\mathrm{ii}+F_\mathrm{ie}$; the $ii$ and $ie$ states for ion-ion Coulomb interactions and electron-screening quantities respectively, and $ee$ is the exchange-correlation contribution in the electron fluid. To describe the mixture excess energy we use
so-called linear mixing rule (LMR) in terms of the free energy of the pure phases
$$f_\mathrm{ex}(Z_{1},Z_{2},\Gamma_{e},x_{1})\approx x_{1}f_\mathrm{ex}(\Gamma_{1},x_{1}=1)+(1-x_{1})f_\mathrm{ex}(\Gamma_{2},x_{1}=0), \eqno(8)$$ where $\Gamma_{i}=\Gamma_{e}Z_{i}^{5/3}$, $x_{1} = N_{1}/(N_{1}+N_{2})$ and $\Gamma_{e}=e^2\beta/a_{e} $ is the electron coupling parameter and $a_{e}=(\frac{4}{3}\pi n_{e})^{-1/3}$ is the mean interelectron distance.
It is worth noting that the calculations in this paper were conducted with several sets of correlation.
For ion-ion interaction from \cite{III} we have
$$f_\mathrm{ii}(\Gamma_{i})=A_{1}(\sqrt{\Gamma_{i}(A_{2}+\Gamma_{i})}-A_{2} ln[\sqrt{\Gamma_{i}/A_{2}}+\sqrt{1+\Gamma_{i}/A_{2}}])+2A_{3}[\sqrt{\Gamma_{i}}-arctan(\sqrt{\Gamma_{i}})], \eqno(9)$$
where $A_{1}=-0.9052$, $A_{2}=0.6322$ and $A_{3}=-(\sqrt{3}/2)-A_{1}/\sqrt{A_{2}}$.
The calculated values of the screening free energy are fitted by the following function from \cite{PtCh}:
$$f_\mathrm{ie}=-\Gamma_{e}\frac{ c_\mathrm{DH} \sqrt{\Gamma_{e}} + c_\mathrm{TF} a \Gamma_{e}^\nu g_{1}(r_{S})}{ 1+[b \sqrt{\Gamma_{e}}+ag_{2}(r_{S})\Gamma_{e}^\nu/r_{S}]}, \eqno(10)$$
$$c_\mathrm{DH}=\frac{Z}{\sqrt{3}}[(1+Z)^{3/2}-1-Z^{3/2}], \eqno(11)$$ $$c_\mathrm{TF} = 0.2513Z^{7/3}(1-Z^{-1/3}+0.2Z^{-1/2}).\eqno(12)$$ where
$a=1.11Z^{0.475},$
$b=0.2+0.078(lnZ)^{2},$
$\nu=1.16+0.08lnZ.$

$$g_{1}(r_{S})=1+\frac{0.78}{21+\Gamma_{e}(Z/r_{S})^3} (\frac{\Gamma_{e}}{Z})^{1/2},\eqno(12)$$
$$g_{2}(r_{S})=1+\frac{Z-1}{9}(1+\frac{1}{0.001Z^2+2\Gamma_{e}})\frac{r_{S}^3}{1+6r_{S}^2}.\eqno(13)$$
The exchange-correlation free energy $f_\mathrm{ee}$ was taken from the \cite{Tanaka}.

General pressure of the system and electrochemical potentials of the ions are obtained from formulas:
$$P_\mathrm{tot} =-(\frac{dF_\mathrm{tot}}{dV})_{N_{i},T},\quad \mu_{i} =(\frac{dF_\mathrm{tot}}{dN_{i}})_{V,T}.\eqno(14)$$

\section {Equilibrium conditions}
Phase equilibrium conditions for two macroscopic phases are well known for the case when coexisting phases consist of arbitrary mixtures of neutral species and include conditions of equilibrium heat and impulse exchange (equality for pressures and temperatures: $P' = P'', T' = T''$) and equations of equilibrium matter exchange\cite{Acta}. They have two variants for systems consisting of two or more chemical elements. The first one is connected with the partial equilibrium for matter with given fixed stoichiometry, or can be equivalent to the well-known Maxwell ``equal squares'' construction for pressure-volume dependence (for example in case of Van der Waals equation of state). More general is the well-known ``double tangent'' construction for two free energies, $F'(V,N_{1},N_{2},T)$ and $F''(V,N_{1},N_{2},T)$ from different equations of state (for example crystal-fluid phase transition). The final equilibrium condition corresponds to equality of Gibbs free energies with fixed chemical composition ($n=N/V$ is the concentration of particles in the system):
$$T' = T'' = T,\quad P'(T, n', x) = P'' (T, n'', x),\quad G'(T, n', x) = G''(T, n'', x).\eqno(15)$$
The second variant corresponds to the equilibrium for exchange by each species with different chemical composition of coexisting phases ($x' \neq x''$), but without changes of total stoichiometry. This variant leads several equalities for chemical potentials of the mixture $\mu_{i}$ ($i=1.2,\ldots,k$--all species), defined as $\frac{\partial F}{\partial N_{i}}_{T,V,N_{j\neq i}}$:
$$P'(T, n', x') = P''(T, n'', x''),\quad \mu_{i}'(T,n', x') = \mu_{i}''(T, n'', x''),\quad \alpha x'+ (1-\alpha)x''= x.\eqno(16)$$
This form is also well known under the name ``Gibbs conditions''. 
The case with equilibrium for charged species is more complicated. The first basic point is that electroneutrality conditions are added for both cases, which were presented above. Maxwell conditions do not change and still valid for Gibbs free energies, $G'$ and $G''$. Besides electroneutrality restrictions two additional quantities appear in description of equilibrium conditions as independent variables. It is average electrostatic potentials, $\varphi'(r)$ and $\varphi''(r)$ \cite{GG}. Therefore, a remarkable feature of any Coulomb system is the existence of two versions of chemical potentials, the ordinary one $\mu_{i}$ (which has to be local quantity depending on local variables) and $\tilde{\mu_{i}}$ (generalized electro-chemical potential, which is not local). The last one depends on non-local sources of influence, such as total charge disbalance including surface dipole, other surface properties etc. In uniform Coulomb system $\tilde{\mu_{i}}$ is equal to the sum of ordinary chemical potentials $\mu_{i}(\left\{n_{k}\right\},T)$ and average electrostatic potential $\varphi = \varphi(r\to \infty)$, which has to be uniform also:
$$\tilde{\mu_{i}} = \mu_{i}(\left\{n(r)\right\},T(r)) + Z_{i}e\varphi(r).\eqno(17)$$ 
Speaking about the equilibrium conditions, the values of chemical potentials must not be equal, but electro-chemical potentials have the same values in coexisting phases:
 $$\mu_{i}'(\left\{n'\right\},T)) \neq \mu_{i}''(\left\{n''\right\},T),\quad \tilde{\mu_{i}'}(\left\{n'\right\},T)) = \tilde{\mu_{i}''}(\left\{n''\right\},T). \eqno(18)$$ 
This form of phase equilibrium equations is known as ``Gibbs-Guggenheim conditions''(GG conditions, c.e.g \cite{Acta,1}).

\begin{figure}[h]
\begin{minipage}{38pc}
\includegraphics[width=38pc]{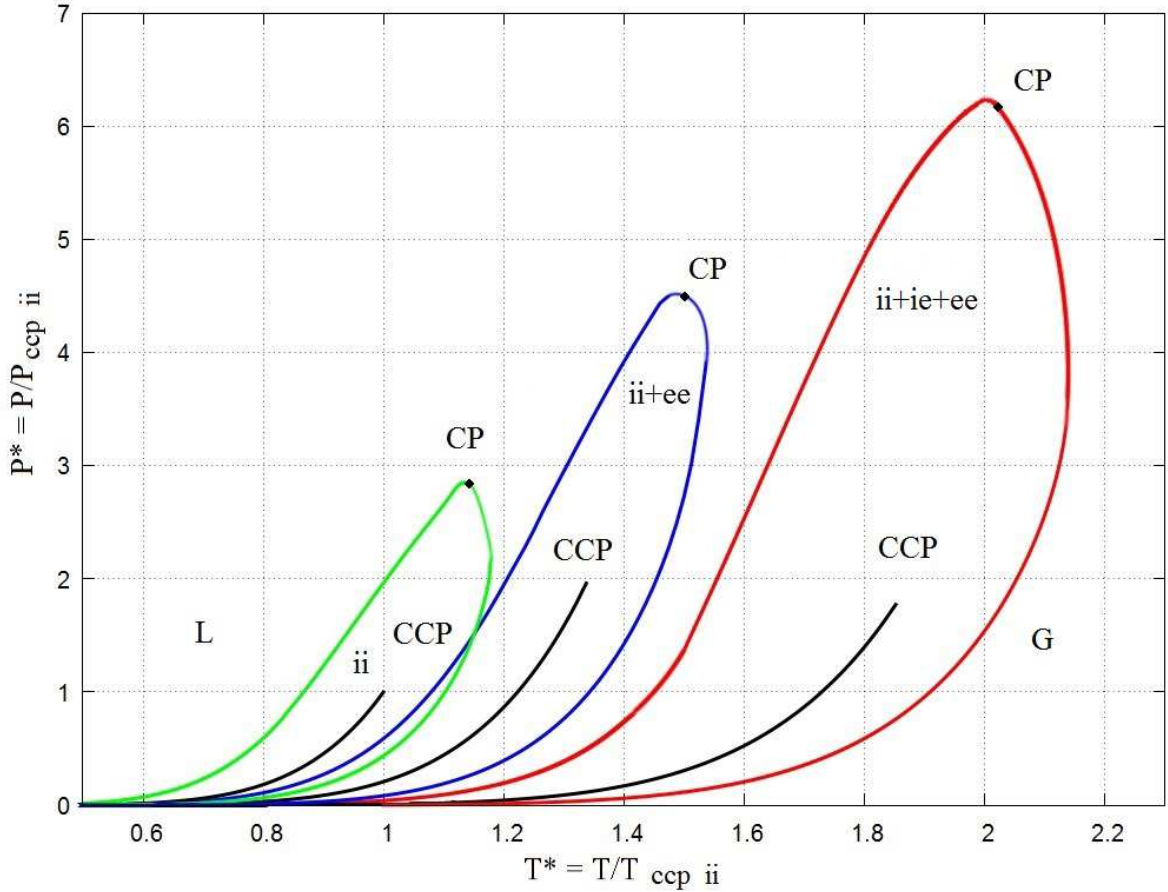}
\caption{\label{label1} $P^*-T^*$ diagrams for carbon-oxygen mixture in congruent (black) and non-congruent equilibrium mode (colored) with marked pseudocritical (congruent critical point --- CCP) and critical points (CP). Green, blue and red lines states for different sets of correlations (see the picture). L and G states for formal liquid and gas phases.}
\end{minipage}\hspace{2pc}%
\end{figure}

\section {Phase diagrams in intensive variables}

In this work the forced-congruent equilibrium (FCE further) regime was used for comparison, and it is that one of the conditions of equilibrium---electrochemical potential equality have to be satisfied for complex bound of ions. In this case we get the so-called model of ``frozen diffusion'', which imposes a strict limit on the change in the composition of the coexisting phases. This model has the typical one-dimensional phase boundary in $P-T$ coordinates for the transition of the gas liquid type and critical point with standard properties ($(\frac{dP}{dV})_{T}=0$ and $(\frac{d^2P}{dV^2})_{T}=0$).

\begin{figure}[h]
\begin{minipage}{38pc}
\includegraphics[width=38pc]{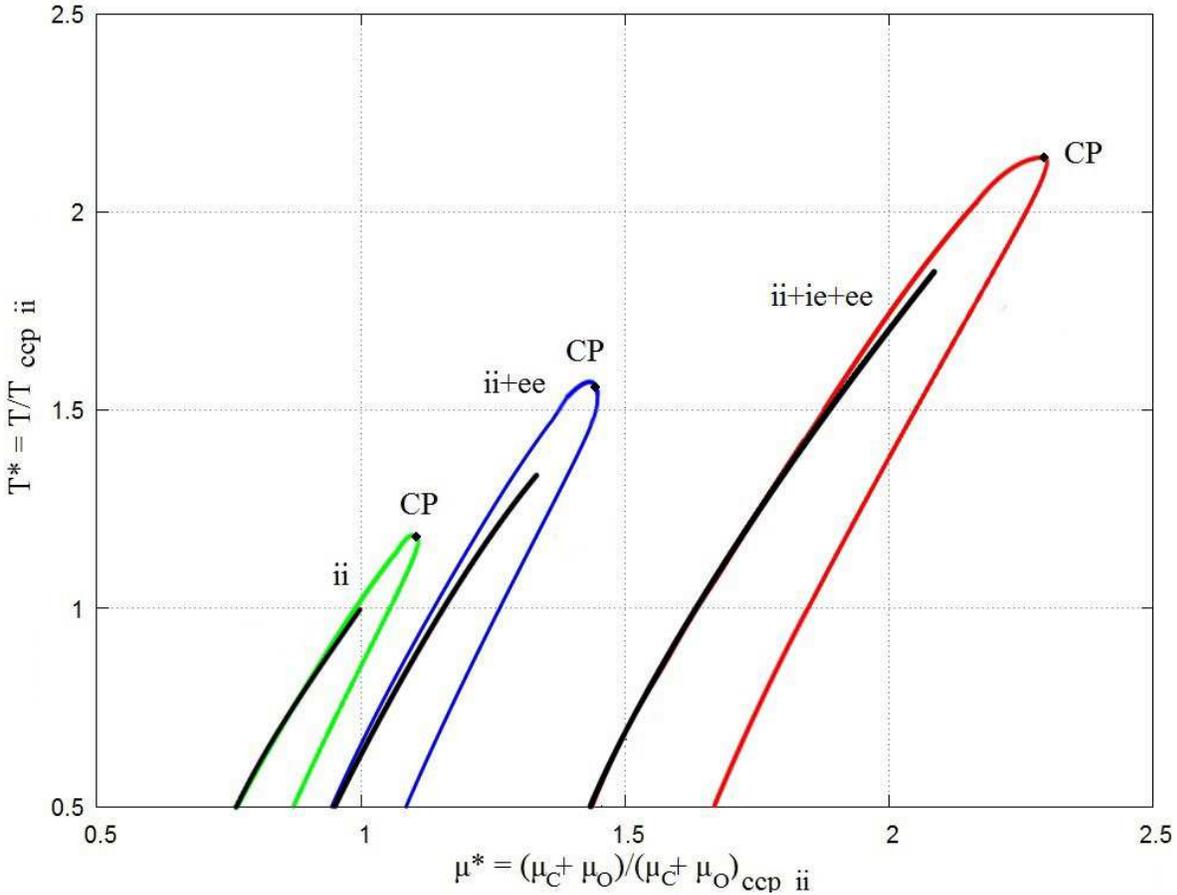}
\caption{\label{label2}$T^*-\mu^*$ diagrams for carbon-oxygen mixture in congruent (black) and non-congruent equilibrium mode (colored). Green, blue and red lines states for different sets of corellations (see the picture).}
\end{minipage} 
\end{figure}

But in general this condition is not required to be satisfied --- instead of a single equality the system of equalities for every electrochemical potential of each component (in this case --- carbon and oxygen ions) have to be satisfied. Then we get non-congruent phase transition, as well as the existence of two solutions instead of one. Above described model with equilibrium conditions allows us to calculate all sets of equilibrium parameters.
One of the solutions corresponds to the start of boiling or the first appearance of vapor bubbles in the liquid phase with a fixed stoichiometry (further BC --- boiling curve, a curve with a fixed ratio in the liquid phase, but with different gas stoichiometry) and another --- to the start of evaporation or the appearance of liquid droplets in a gas (further SC --- saturation curve, fixed ratio in the gas phase, different liquid stoichiometry). This leads to the appearing of two-phase zone in intensive variables in contrast to the FCE mode. It should be noted that the liquid and gas are the formal designation of the phases with the highest density (and therefore the degree of interaction) and the least respectively, because of the fully ionized plasma case.

The same ``banana-like'' structure of 2-phase region was obtained in research of non-congruent phase transition in the uranium dioxide model \cite{IoYakub,UO2} and in the study of liquid-gas phase transition in nuclear matter \cite{PRC}.
BC and SC solutions join in the critical point (``true '' critical point or CP further), that does not match with the point obtained in FCE mode (congruent critical point or CCP). The pictures shows the phase diagram in the intensive variables for the three models, different sets of interparticle correlation (see figures~1 and 2).

 The simplest model is located in a low-temperature region and describes an ideal gas of ions on the background of the ideal electrons taking into account only the ion-ion correlations. 
Adding new non-ideal part moves critical (as well as pseudocritical) point to higher temperatures, the following model takes into account electron-electron interaction and corresponds to DOCP (double OCP) model in case of one-component plasma (mixed OCP of ions and OCP of electrons excluding the interparticle interactions between these varieties) and last---full EoS of Potekhin and Chabrier with all interactions.

\begin{figure}[h]
\begin{minipage}{17pc}
\includegraphics[width=17pc]{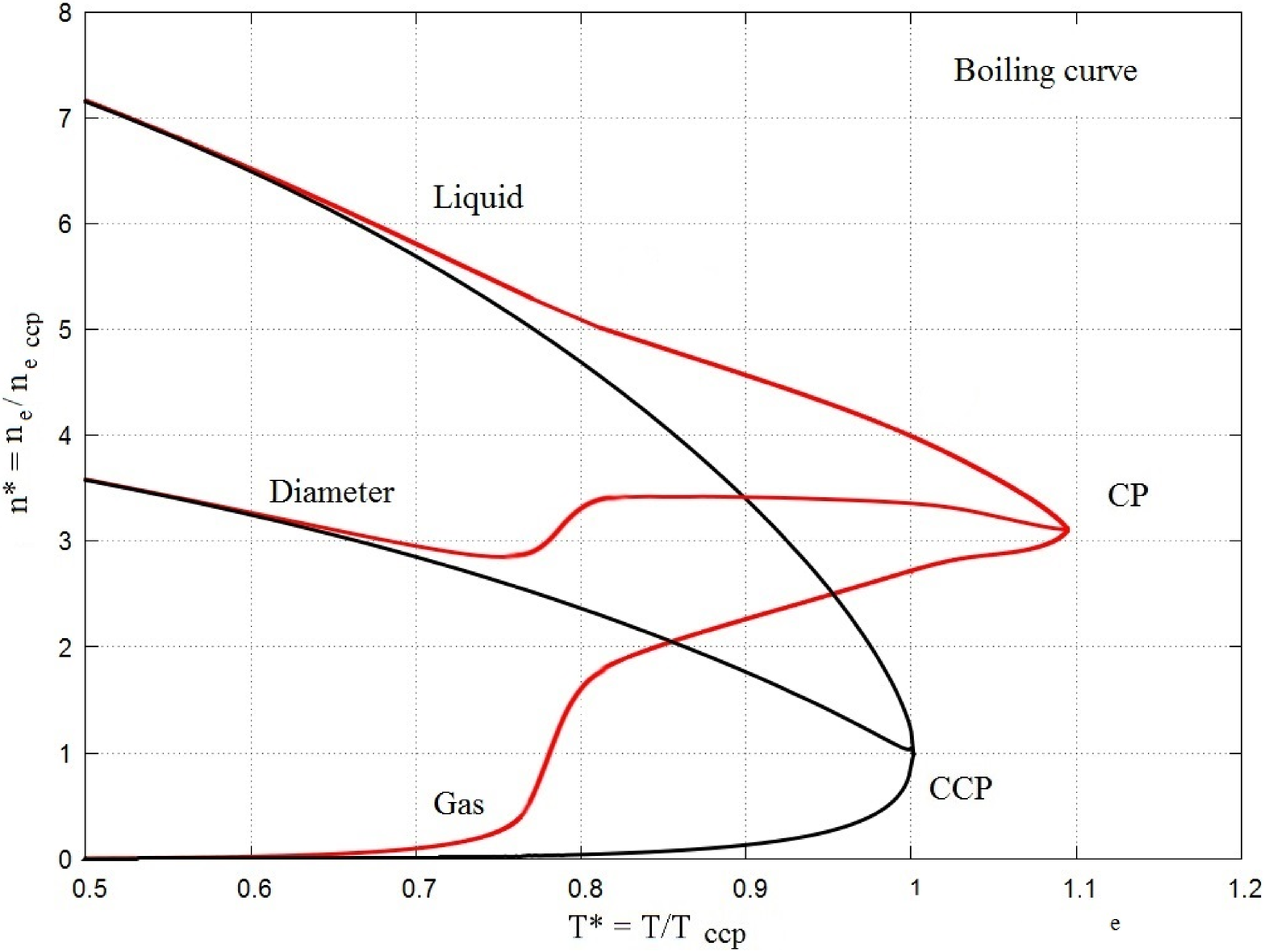}
\caption{\label{label3}$n_\mathrm{e}^*-T^*$ (Electron concentrations (density analogue $n_{e}=Z_{1}n_{1}+Z_{2}n_{2}$)---normalised temperature) for equilibrium according to the calculations with full Potekhin and Chabrier EoS. Black lines---congruent mode of equilibrium, red---non-congruent. Upper lines state for liquid phase, lower---gas, middle line---diameter(half-sum of the densities for coexisting gas and liquid phases). The picture describes the boiling curve.}
\end{minipage}\hspace{2pc}%
\begin{minipage}{17pc}
\includegraphics[width=17pc]{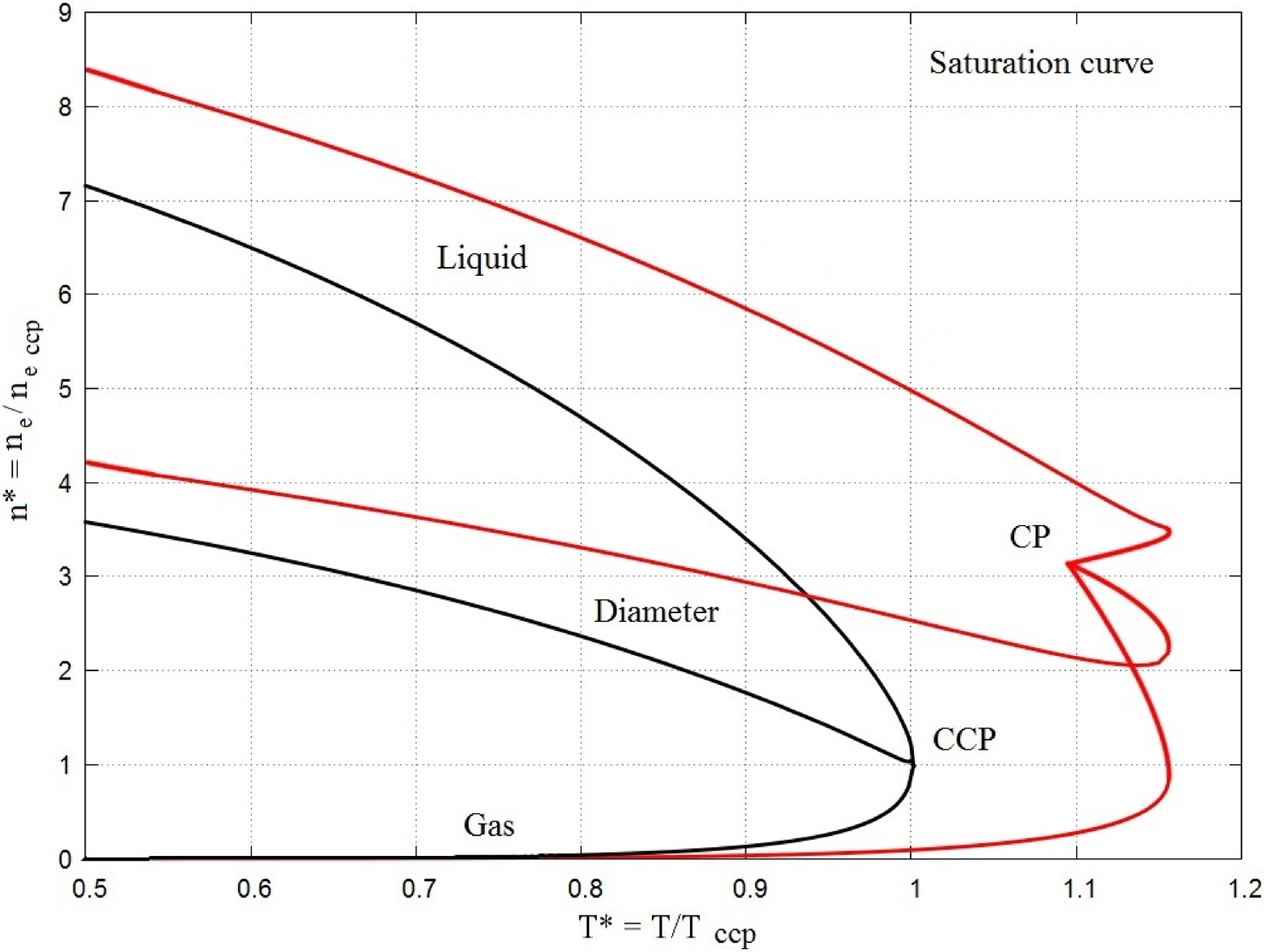}
\caption{\label{label4}$n_\mathrm{e}^*-T^*$ (Electron concentrations (density analogue $n_{e}=Z_{1}n_{1}+Z_{2}n_{2}$)---normalised temperature) for equilibrium according to the calculations with full Potekhin and Chabrier EoS. Black lines---congruent mode of equilibrium, red---non-congruent. Upper lines state for liquid phase, lower---gas, middle line---diameter(half-sum of the densities for coexisting gas and liquid phases). The picture describes the saturation curve.}
\end{minipage} 
\end{figure}

The studied model allows us to consider the majority of the thermodynamic quantities and parameters responsible for balance in the system. 
Phase diagrams for the concentration of particles in the system depending on temperature are built for full EoS (see figures~3 and 4). The behavior of the phase boundaries in the congruent and non-congruent mode of equilibrium differs significantly from each other. The first noticeable difference is the coexistence of two scenarios---BC and SC (see above), the last is remarkable for its retrograde area, which also has several coexisting phases at fixed temperature.
Another notable difference is a very different position of the critical point, unlike in the FCE regime and other properties of CP (which will not be examined in details in this work).

In the study of real substances and Coulomb models, there are many semi-empirical rules that allow to make accurate assessments about the thermodynamic parameters of the system, including the properties and position of the critical point. Traditional rule of ``rectilinear diameter'' of Calliete Matthias for density-temperature two-phase boundary is well known for many years. According to this semi-empirical rule the dependence of half-sum of the densities (``the diameter'') for coexisting gas and liquid phases on temperature is almost linear usually up to close vicinity of critical point. This is true for congruent mode, but in non-congruent case the diameter does not lead to the CP and also difficult to be approximated with a straight line.

\section {Galvani potential}
Every phase boundary in Coulomb system is accompanied by a finite gap $\Delta \varphi$ in the average electrostatic potential through the phase interface \cite{IoFizmatlit,Acta}, because of the GG equalities combined with the electroneutrality conditions.
It is worth mentioning that because of the last condition there is no charges on the interphase border.

\begin{figure}[h]
\begin{minipage}{17pc}
\includegraphics[width=17pc]{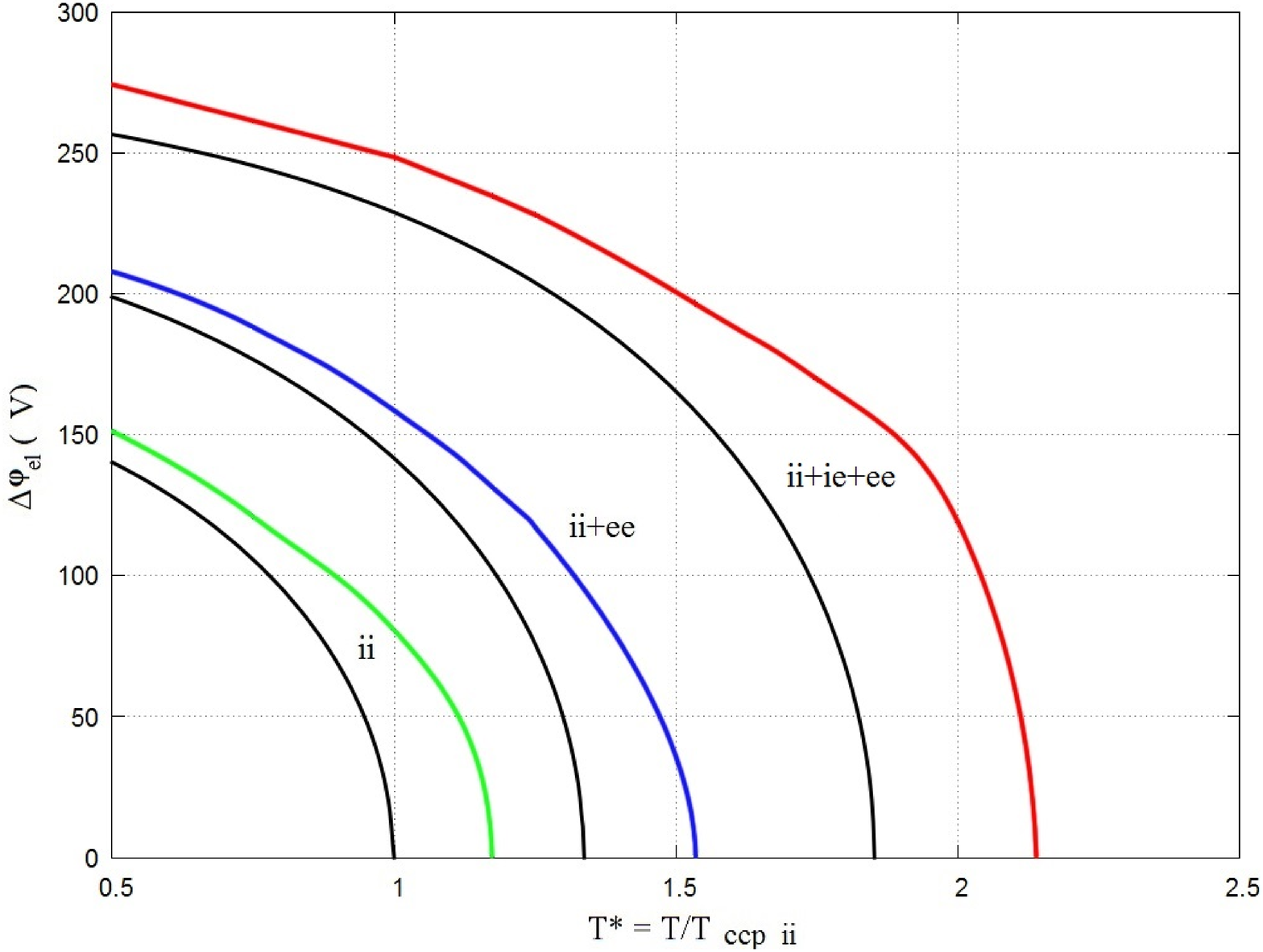}
\caption{\label{label5}$\Delta\varphi (V)-T*$ diagram: Electrostatic potential of gas-liquid interface (Galvani potential) in congruent (black) and non-congruent (colored) scenario of phase coexistence in equimolar C-O mixture (1:1). Correlations sets are given on the picture.}
\end{minipage}\hspace{2pc}%
\begin{minipage}{17pc}
\includegraphics[width=17pc]{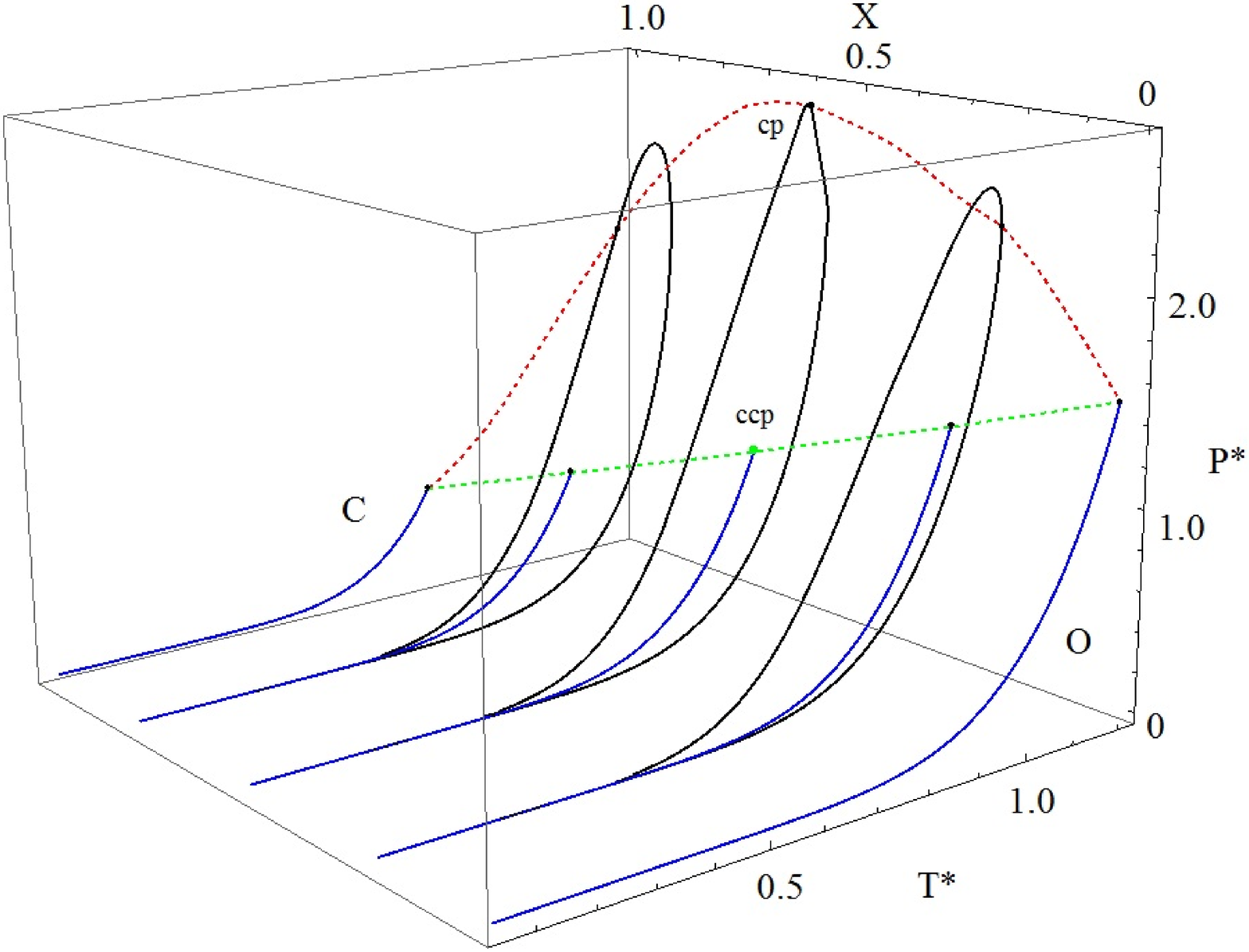}
\caption{\label{label6}$P^*-T^*-X$ diagram for fully ionized plasma (only with ion-ion corellations) with different C-O composition and two versions of critical points line. Blue lines---$P^*(T^*)$ for pure C and O plasmas, for C-O (X=0.5) mixture, C-O (X=0.23) and C-O (X=0.78) in forced-congruent regime; Black lines---the same for mixtures in case of non-congruence with 2 solutions of equilibrium; Green dashed line consists of congruent critical points (or pseudocritical points); Red dashed--the set of true critical points. }
\end{minipage} 
\end{figure}

$\Delta \varphi$ states for the thermodynamic quantity, which is influenced only by temperature and chemical composition and not surface properties. 
The equilibrium potentials of electrons were explicitly taken from the calculated parameters and according to its difference the Galvani potentials were built for different models in this work.
This difference with increasing temperature decreases monotonically to the critical point, but the magnitude of the jump depends on the model and the regime of equilibrium (see figure~5).

\section {Additional features of the BIM($\sim$)}
The so-called ``end'' points---with a maximum pressure (PMP---pressure maximum point) and the maximum temperature (TMP) appear in the two-dimensional coordinates, and in general they do not coincide with the CP (except certain ratios). The difference in the locations of these points can be observed in \cite{StIo-15}.
Parameters of critical points line were calculated for the entire range of proportions of ions, as well as the parameters of this line in congruent mode. There is a significant difference in the behavior of these curves, most notably manifested at an equal ratio of components (see figure~5).

By observing the concentrations of ions behavior (and the total number of particles because of the electroneutrality condition) depending on the temperature in different solutions (BC and SC) the distillation effect was observed. It influences the equilibrium stoichiometry, the system advantageously consists of mainly the one component. From the thermodynamics point of view and minimizing Helmholtz Free energy system moves to the one component in the coexisting phase. The graphs can be seen in work \cite{StIo-15}.

This distillation effect can be also observed in nuclear matter, which was mentioned above \cite{PRC}, while at the same time for comparison can be demonstrated in \cite{IoYakub,UO2} for chemically active plasma, where this effect is not present, and the phenomenon of non-congruence drops at low temperatures. 

\section {Summary}
Non-congruent gas-liquid phase transition have been studied in modified Coulomb model of a binary ionic mixture /BIM($\sim$)/. The studied model allowed us to calculate the majority of the thermodynamic equilibrium values and to build different phase diagrams. In this paper the comparison of the obtained calculation results for models with different sets of nonidealities was demonstrated. Parameters of critical points-line was calculated on the entire range of proportions of mixed ions $0<X<1$. Strong ``distillation'' effect was found for NCPT in present BIM($\sim$). The Galvani potential was calculated also, which plays an important role in the equilibrium of the Coulomb system. Several papers were mentioned for comparison of the results obtained and the features of the model.

\section {Acknowledgement}
This work was supported by the Russian Science Foundation, grant 14-50-00124.

\section*{References}
\bibliography{iopart-num}

\end{document}